# Breaking time reversal symmetry in topological insulators

Cui-Zu Chang[1]*, Peng Wei,[1] Jagadeesh. S. Moodera[1,2]*
[1]Francis Bitter Magnetic Lab, Massachusetts Institute of Technology, MA02139, USA
[2]Department of Physics, Massachusetts Institute of Technology, MA02139, USA

**A wide class of materials that were discovered to carry a topologically protected phase order has led to a highly active area of research called topological insulators (TIs). This phenomenon has radically changed our thinking because of their robust quantum coherent behavior showing two-dimensional Dirac-type metallic surface states (SSs) and simultaneously insulating bulk states. The Dirac SSs are induced by the strong spin–orbit coupling (SOC) as well as protected by the time reversal symmetry (TRS). Breaking TRS in a TI with ferromagnetic perturbation can lead to many exotic quantum phenomena, such as the quantum anomalous Hall effect, topological magnetoelectric effect, as well as image magnetic monopole. This article presents an overview of the current status of TRS breaking in TIs and outlines the prospects for future studies.**

**Introduction**

Topological insulators (TIs), predicted and observed, display helical metallic surface states (SSs) obeying a Dirac-type linear energy-momentum dispersion relation that are protected by time reversal symmetry (TRS).[1–6] Upon a time reversal operation, which lets the system to evolve backward in time, the electron wave vector *k* and the spin will flip the sign. The helical surface states of a topological insulator are invariant under such operation since the opposite spin channels are locked to the opposite momentums. In the presence of magnetic field or magnetic impurities, however, this invariant or symmetry will be broken. Although the TI is an ordered phase not relying on broken symmetry, the symmetry broken states created in a TI have been predicted to carry many novel quantum phenomena[1,2] (e.g., the quantum anomalous Hall effect [QAHE],[7–8,9] topological magnetoelectric effect,[7,10] as well as image magnetic monopole[11]). The unique properties exhibited by these exotic systems open up avenues for both fundamental physics research and new materials displaying exotic phenomena aimed at technological applications.

The spontaneously broken time reversal symmetry states can be experimentally introduced into a material by ferromagnetic ordering. This may usually be achieved by two methods that are described in the following paragraphs: (1) conventionally by doping with some magnetic element and (2) by ferromagnetic proximity coupling. In both cases, it is expected that an exchange gap opens in the Dirac SS.[7] Although this seems straightforward, the major difficulty remains in reducing the bulk conduction, particularly in thin films, and thus it is a daunting task for any TI material system. This difficulty hinders observations of the predicted properties of TIs with TRS broken.

Similar to conventional diluted magnetic semiconductors (DMSs),[12–14] impurity doping using transition metal (TM) elements (e.g., Cr, V, Mn) is a convenient approach to induce long-range ferromagnetic order in TIs (**Figure 1**a). Many recent experiments, from angle-resolved photoemission spectroscopy (ARPES) to electrical transport measurements, have been devoted to the study of magnetically doped TIs of the $Bi_2Se_3$ family, including $Bi_2Se_3$, $Bi_2Te_3$, and $Sb_2Te_3$.[15–23] In magnetically doped $Bi_2Te_3$ and $Sb_2Te_3$, both transport and magnetization measurements have shown clear long-range ferromagnetic order in several cases,[20–23] which resulted in the observation of QAHE (the quantum Hall effect [QHE] without an external magnetic field).[23] The Zeeman gap opening, which is the splitting of the energies between the spin-up and spin-down electron states under an external or effective internal magnetic fields, at the SS Dirac cone, however, has not been resolved by ARPES or scanning tunneling spectroscopy (STS) in ferromagnetic TIs, possibly due to their low Curie temperature ($T_C$) and small gap size. On the other hand, for magnetically doped $Bi_2Se_3$, the SS Zeeman gap has been reported in several ARPES studies, with gap size varying from several tens to a hundred meV, implying strong ferromagnetism.[15,16] However, transport and magnetization measurements are only able to show very small or no magnetic hysteresis in the perpendicular magnetic field, hindering further progress toward observing QAHE in Se-based TIs.[17,21]

Proximity-induced ferromagnetism in a TI is attainable in a heterostructure of TI and a ferromagnet obtained by utilizing a thin-film deposition technique. Commonly used ferromagnetic layers are Fe, Co, and Ni, which are metallic. However, they do not grow very well on top of a TI surface due to the low surface energy and strong oxidizing nature of tellurium and selenium of the TI material.[4,5] Most importantly, as a metallic thin-film layer parallel to a TI, these ferromagnetic materials overwhelm the TI conduction. Magnetic adatoms, such as Fe or Mn, may be used to avoid forming a continuous conducting layer,[24,25] whereas these adatoms acting as magnetic impurities can lead to spin scattering, and thus is detrimental to TI SSs.[26] Therefore inert and insulating ferromagnetic materials (i.e., ferromagnetic insulators [FIs]), such as EuS, would be highly advantageous.[7] The short-range nature of magnetic proximity coupling with an FI allows TI SSs to experience ferromagnetic interactions. In this case, the all-important advantage is that the TRS breaking happens mostly at the interface between the TI and the FI, rather than affecting the majority bulk states, while also not introducing defects (Figure 1b). However, in general, thin films of FIs show in-plane easy axis of magnetization,[27–29] which hinders the observation of the QAHE or the zero-field half-integer QHE.[7] We provide an overview of the behavior of FI/TI heterostructures and that of magnetically doped Se/Te-based TI compounds as well as bring out their differences. We also describe the experimental observation of QAHE in Cr-doped $(Bi_xSb_{1-x})_2Te_3$ thin films.

**FI/TI heterostructures**

Creating a heterostructure of a 3D TI and a FI is of great interest. Several phenomena, predicted for TIs with broken TRS, exist only at the interface. Contrary to the QAHE of magnetically doped TI films,[23] a half-integer QAHE can be observed only in a FI/TI heterostructure.[7] Furthermore, a FI layer introduces TRS locally, and thus a lithographically patterned FI thin film can open a Zeeman gap at selected regions on a TI surface, an essential prerequisite for confining Majorana fermions (fermions that are their own antiparticles) in a superconducting TI.[30,31] One of the excellent properties exhibited by adjoining the FI to a superconducting TI layer is that the local exchange field experienced by the TI is able to lift spin degeneracy without destroying the superconducting pairing, in complete contrast to magnetic impurity doping.[32,33] This opens up significant new avenues for functional ferromagnetic superconducting TI devices. One of the most critical factors in this proximity-induced approach is interfacial control of the TI/FI system: not only should they be closely interacting magnetically, but they should also be maintaining a chemically 'intact' interface; control down to a single atomic layer is necessary to achieve this.

FI/TI heterostructures have shown success in producing ferromagnetism in TIs by the proximity effect with the choice of EuS/$Bi_2Se_3$.[27] An induced interface magnetization was confirmed by various observations. The thickness dependence of the magnetic moment in EuS/$Bi_2Se_3$ bi-layers measured with a SQUID magnetometer exhibited excess moment coming from the ferromagnetic TI surface.[27] The ferromagnetic order in TI was further evident from the anomalous Hall effect in the bi-layers.[27] With similar objectives, Yang et al.[28] and Kandala et al.[29] investigated the effects on the weak anti-localization (WAL) behavior of the TI in conjunction with an FI, where the electrons with spin-momentum locking are less likely to encounter backscattering hence become localized once the time reversal symmetry is preserved. The planar magnetoresistance (MR) of these heterostructures, studied using patterned Hall bars with in-plane applied magnetic field (EuS has an in-plane easy axis), showed hysteretic MR signifying proximity-induced ferromagnetic order of charge carriers in $Bi_2Se_3$.[27] **Figure 2**a shows an example of this for a 1 nm EuS/20 nm $Bi_2Se_3$ bi-layer. The sample resistance undergoes a minimum at the bi-layer's ferromagnetic coercive field ($H_C$), independent of the angle between the applied in-plane field and the electric current. The large number of random magnetic domains that develops (Figure 2b) at $H_C$ coincides with the occurrence of resistance minima (or higher electrical conduction). This unconventional behavior was attributed to the isotropic distribution of domains and hence the observed isotropic planar MR. Interestingly, the Dirac SSs along the domain wall were theoretically predicted to have zero mass carrying chiral conduction modes similar to the quantum Hall edge channels,[1,2] in agreement with the experimental observations of the MR dips (increased conduction) at $H_C$.[27]

Generally, all the FIs have an in-plane easy magnetization axis, although canted magnetization or small perpendicular magnetic hysteresis has been reported.[27–29] However, the perpendicular magnetic anisotropy (PMA) can be engineered by choosing a proper interface layer, which is attributed to the hybridization of the electronic states.[34] Strong PMA in transition metal/rare-earth multilayers has been reported.[35,36] Thus the proper choice of a FI and TI material combination may lead to a strong PMA. Nevertheless, it may be noted that QAHE has also been predicted in FI/TI heterostructures with an in-plane easy magnetization axis, although this is yet to be experimentally confirmed.[37]

**Magnetically doped TI materials**

Doping 3d TM impurities into TIs is a convenient approach to bring robust long-range ferromagnetic order, as demonstrated by the success of diluted magnetic semiconductors (DMSs).[12–14] Doping TI compounds with magnetic atoms could be a straightforward way to reach the goal. However, it is important to confirm the absence of other secondary phase formation in the bulk or locally on the surface, and/or clustering or non-uniform distribution of dopants in the parent compound (all of which can yield a false magnetic signature). The results become fortuitous and unreliable, unless the above factors are carefully addressed to rule them out.

Ferromagnetism has been realized in several TI materials of the $Bi_2Se_3$ family doped with 3d TM ions.[15–23] The observation of a Zeeman gap at the SS Dirac cone, a sign of broken TRS, has been reported in Fe- or Mn-doped $Bi_2Se_3$.[15,16] **Figure 3** shows one example of ferromagnetic TI with Cr doping; the magnetic moment per Cr ion was observed to be close to $2\mu_B$, with an optimum hysteresis loop for a Cr content of 5.2%.[17] For Cr concentrations above or below this, hysteresis was present at lower temperatures. The participation and magnetic nature of charge carriers in the doped TI was seen in the planar MR signal (inset of Figure 3), while in a perpendicular magnetic field, no hysteresis was seen (preferentially in-plane magnetization).[17] The Hall traces exhibit a linear dependence on the perpendicular magnetic field, and the MR exhibited WAL behavior at 10 K, both of which are the characteristics of a TI with TRS. Below 10 K, a nonlinear Hall signal gradually developed, accompanied by the evolution of MR into weak localization behavior, which implies broken TRS. However, long-range ferromagnetic order with perpendicular easy axis is absent since no anomalous Hall loop was observed.[18,19] Comparing the in- and out-of- plane magnetization data, Cr-doped $Bi_2Se_3$ thin films displayed in-plane magnetization.

A well-defined anomalous Hall effect (AHE), however, was seen in Mn-doped $Bi_2Te_3$,[20] and Cr-doped $Bi_2Te_3$ or $Sb_2Te_3$ films.[21–23] To understand the contrasting magneto transport properties of Cr-doped $Bi_2Se_3$ and $Bi_2Te_3$, Zhang et. al. studied 8 quintuple layers (QL) Cr-doped $Bi_2(Se_xTe_{1-x})_3$, an isostructural, isovalent mixture of $Bi_2Se_3$ and $Bi_2Te_3$. The Hall traces of all the films measured at $T = 1.5$ K are summarized in **Figure 4**a, revealing a highly systematic evolution of coercivity ($H_C$) and the AHE resistivity $\rho_{yx}^0$ at zero applied magnetic field. They further constructed a magnetic phase diagram by plotting the $\rho_{yx}^0$ values as a function of the Se (or Te) concentration. At the base temperature of $T = 1.5$ K, the phase diagram was separated into two distinct regimes: a ferromagnetic phase with positive $\rho_{yx}^0$ and a paramagnetic phase with negative $\rho_{yx}^0$, as shown in Figure 4b. Because the transition between the two magnetic phases occurs at the ground state, it is a quantum phase transition driven by the change of chemical composition, as confirmed by ARPES measurements and density functional theory calculations.[21]

**Quantum anomalous Hall effect**

The quantized version of AHE, QAHE, predicted to be one of the representative quantum phenomena displayed by TRS-broken TIs, was observed experimentally in a system with five QL, Cr-doped $(Bi_xSb_{1-x})_2Te_3$ thin films.[23] In order to observe QAHE in a TI, one needs three prerequisites:[7] First, the growth of TI quantum well thin films on insulating substrates (e.g., sapphire (0001) or $SrTiO_3$ (111)).[38] Thinner TI films promote localization of dissipative channels and result in a gap between the lowest order quantum well subbands, larger than the ferromagnetic exchange gap. Second, a long-range ferromagnetic state in the TI that should continue into the insulating regime and have an easy magnetization axis perpendicular to the film plane is another necessity. Third, the chemical potential should be tunable into the magnetic gap. As discussed in the previous section, ferromagnetic Cr-doped $Bi_2Se_3$ has in-plane magnetization. On the other hand, Mn-doped $Bi_2Te_3$, Cr-doped $Bi_2Te_3$, and $Sb_2Te_3$ show perpendicular magnetic anisotropy. Among these materials, Cr-doped $Sb_2Te_3$ seemed to be the best, as it showed carrier-type and density-independent AHE, hence the QAHE.[23]

The $p$-type conduction in $Sb_2Te_3$ is usually due to anti-site defects,[39] whereas $Bi_2Te_3$ is intrinsically $n$-type and has the same structure as $Sb_2Te_3$.[40] Thus, by partially replacing Sb with Bi in Cr-doped $Sb_2Te_3$, the chemical potential is tunable across the Dirac point as was confirmed by ARPES measurements.[22] **Figure 5** shows the magnetic field dependent Hall resistance of 5QL $Cr_{0.22}(Bi_xSb_{1-x})_{1.78}Te_3$ films grown on sapphire (0001) substrates with different Bi:Sb ratios keeping the Cr doping a constant. At $T = 1.5$ K, all films showed near square-shaped AHE hysteresis loops, suggesting strong ferromagnetism with perpendicular axis magnetization. With increasing Bi concentration, the ordinary Hall effect evolved from positive to negative, showing carrier type changing from $p$- to $n$-type, with no effect on the ferromagnetism in Cr-doped $(Bi_xSb_{1-x})_2Te_3$ despite a significant change of the carrier density as well. The Curie temperature ($T_C$) stayed unchanged, even in the rather insulating samples around the $p$- to $n$-transition. This showed carrier independent ferromagnetism, supporting the existence of a FI phase likely by a van Vleck mechanism.[7] The anomalous Hall (AH) resistance, on the other hand, exhibited a dramatic and systematic change with Bi doping. The largest AH resistance was ~3 kΩ, when the sample had the lowest carrier density. Although this number is larger than the AH resistance observed in most ferromagnetic metals, it still is far from the quantized value (~25.8 kΩ).[22]

Further optimization of sample growth and measurements at ultralow temperature (~30 mK) yielded QAHE in 5QL Cr-doped $(Bi_xSb_{1-x})_2Te_3$ thin films on $SrTiO_3(111)$ substrates. The high dielectric constant of $SrTiO_3$ at low temperature allowed for the possibility of applying a sufficiently large electric field to fine tune the carrier density. **Figure 6**a–b shows the magnetic field dependence of the Hall resistance ($\rho_{yx}$) and longitudinal resistance ($\rho_{xx}$), respectively, measured at T = 30 mK for various gate bias ($V_g$). The shape and coercivity of the hysteresis loops are nearly unchanged, with $V_g$ showing carrier independent ferromagnetism.[7] At the saturated magnetization state, $\rho_{yx}$ is nearly independent of the magnetic field, suggesting uniform ferromagnetic order and charge neutrality of the sample, whereas the AH resistance changed dramatically with $V_g$, with a maximum value of $h/e^2$ (25.8 kΩ) occurring at $V_g$ ~1.5 V. The MR curve exhibited the typical hysteretic shape as in ferromagnetic metals.[23]

The most important observation, the Hall resistance exhibiting a distinct plateau with the quantized value $h/e^2$ at zero applied magnetic field (QAHE), is shown in Figure 6c. Complementary to this $\rho_{yx}$ plateau, the longitudinal resistance $\rho_{xx}(0)$ shows a dip, reaching a value of 0.098 $h/e^2$, yielding a $\rho_{yx}(0)/\rho_{xx}(0)$ ratio (Hall angle) of 84.4°. These results demonstrated the realization of QAHE in magnetically doped TI films. It may be noted that, compared with QHE systems, all of these doped TI samples reported/displaying QAHE had a rather low mobility (<1000 cm$^2$/Vs). The observation of QAHE not only shows the ability to obtain QHE without Landau levels but also paves a way for developing low-power future electronics. It also brings out the feasibility for discovering other predicted exotic quantum phenomena in TIs, such as the topological magnetoelectric effect[7] and Majorana bound states.[30,31]

**Conclusions and prospects**

We have provided a brief review of the phenomenon of broken TRS in TIs, which is achievable, both through the proximity effect in FI/TI heterostructures and magnetic impurity doping. The proximity-induced ferromagnetism originates from a uniform interface exchange field that does not structurally disturb the TI. In transition metal ions-doped TIs, long-range ferromagnetic order was found in Cr (or Mn)-doped Se- and Te-based TIs. In carefully optimized and electric field tuned Cr-doped (Bi$_x$Sb$_{1−x}$)$_2$Te$_3$ thin films, QAHE was demonstrated, while the ferromagnetism in the system was seen to be insensitive to the carrier type and density, demonstrating a novel magnetic TI phase.

The proximity effect of a FI/TI heterostructure not only enables the creation of cleaner ferromagnetic TIs with larger surface exchange gap, but also allows locally breaking TRS in designated areas over a TI surface by lithographic patterning. This might be used to create exotic domain wall states between the TI surface regions with and without TRS. Furthermore, a FI can form a seamless interface with a topological superconductor to confine Majorana bound states by a proximity-induced Zeeman gap, whereas magnetic impurities may easily destroy the superconducting pairing.[30–32] Besides, the zero-field half integer quantum Hall effect is also expected to be a hallmark of the FI/TI/FI sandwich structure.[7] In magnetic element-doped and ferromagnetic TIs, the observation of QAHE has set the possibility and hope for observing many other exotic quantum phenomena predicted in TIs, such as a topological magnetoelectric effect,[7] image magnetic monopoles,[11] and Majorana bound states.[30,31] Importantly, since the QAHE inherits high mobility quantum Hall edge channels, low-power-consumption spintronic devices will be significantly promoted.

Contact information:
Cui-Zu Chang, email: czchang@mit.edu;
Peng Wei, email: pwei@mit.edu;
J. S. Moodera, email: moodera@mit.edu.

**Acknowledgments**

C. Z. C. would like to acknowledge collaborations with Q.K. Xue, K. He, and Y.Y. Wang at Tsinghua University, China. P.W. and J.S.M. would like to acknowledge collaborations with D. Heiman and P. Jarillo-Herrero. We thank funding support from the MIT MRSEC under the NSF Grant DMR-0819762, the NSF Grant DMR-1207469, the ONR Grant N00014–13–1–0301, and the STC Center for Integrated Quantum Materials under NSF Grant DMR-1231319.

# Figure Captions

**Figure 1.** Sample structures of (a) magnetically doped topological insulator films and (b) a ferromagnetic insulator (FI)/topological insulator (TI) heterostructure.

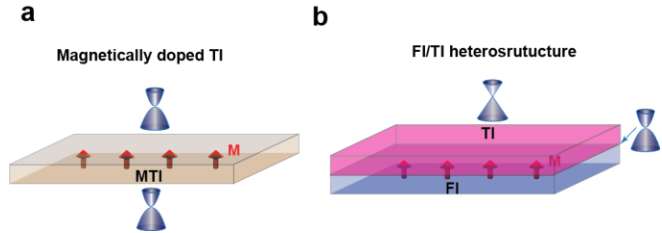

**Figure 2.** (a) Planar magnetoresistance of a $Bi_2Se_3$/EuS sample measured at $T = 1.0$ K indicates domain wall-assisted electron conduction. Curves are shifted along the y-axis for clarity. (b) Schematic diagram showing the inferred magnetic moment distribution between magnetic domains at the $Bi_2Se_3$/EuS interface. Figures adapted from Reference 27.

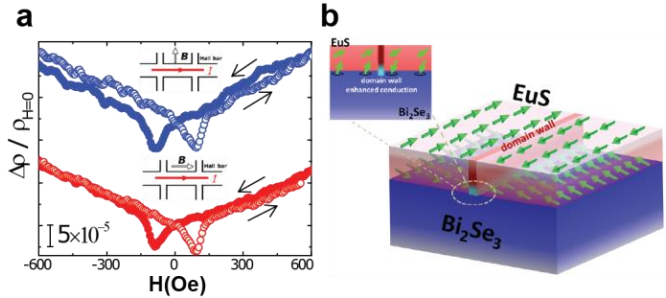

**Figure 3.** Magnetic moment per Cr ion as a function of applied in-plane magnetic field at $T = 4$ K. Inset: planar magnetoresistance signal in a sample with $x = 1.3\%$ at $T = 1$ K, showing clear hysteresis. Adapted from Reference 17.

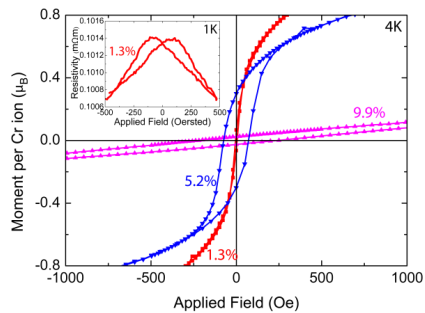

**Figure 4.** (a) Systematic evolution of the Hall effect for all the samples ($0 \leq x \leq 1$) in 8 QL $Bi_{1.78}Cr_{0.22}(Se_xTe_{1-x})_3$ measured at $T = 1.5$ K. (b) Magnetic phase diagram of $Bi_{1.78}Cr_{0.22}(Se_xTe_{1-x})_3$ summarizing the intercept $\rho^0_{yx}$ as a function of $x$ and $T$. The $T_C$ of the ferromagnetic phase (FM) is indicated by the solid symbols (PM = paramagnetic phase). Adapted from Reference 21.

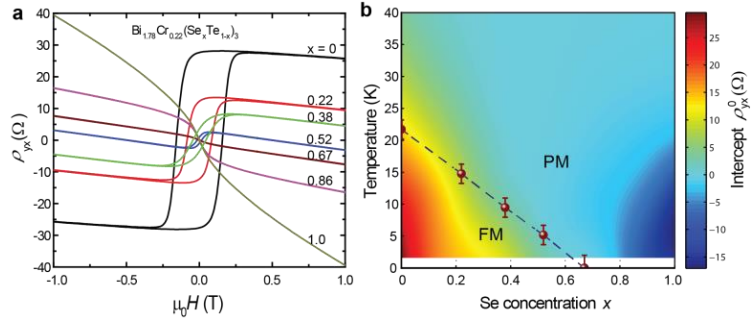

**Figure 5.** Magnetic and electric properties of Cr-doped $(Bi_xSb_{1-x})_2Te_3$ films. Magnetic field dependent Hall resistance $R_{yx}$ of the $Cr_{0.22}(Bi_xSb_{1-x})_{1.78}Te_3$ films with (a) $x = 0$, (b) $x = 0.15$, (c) $x = 0.2$, (d) $x = 0.25$, (e) $x = 0.35$, and (f) $x = 0.5$ at various temperatures. Adapted from Reference 22.

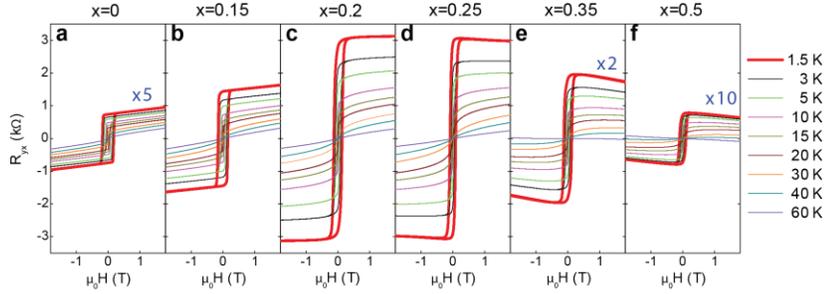

**Figure 6.** The quantum anomalous Hall effect measured at 30 mK. (a) Magnetic field dependence of $\rho_{yx}$ at different $V_g$ values. (b) Magnetic field dependence of $\rho_{xx}$ at different $V_g$ values. (c) Dependence of $\rho_{yx}(0)$ (empty blue squares) and $\rho_{xx}(0)$ (empty red circles) on $V_g$. The vertical purple dash-dotted lines in (c) indicate $V_g$ for the charge neutral point ($V_g^0$). Figure adapted from Reference 23.

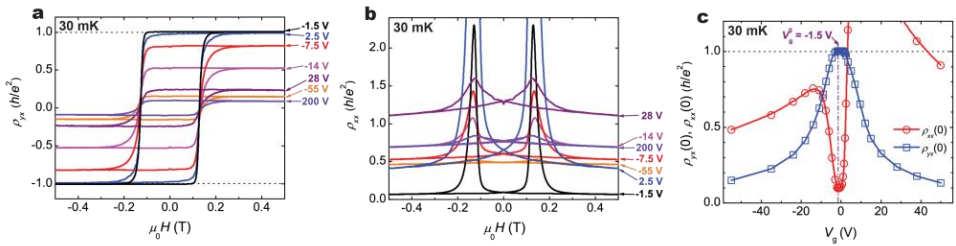